# Descriptor: Five years of meteorological surface data at Oak Ridge Reserve in Tennessee

**MORGAN R. STECKLER[1], KEVIN R. BIRDWELL[1], HAOWEN XU[2], AND XIAO-YING YU[3,*]**

[1]Integrated Computational Earth Sciences Group, Oak Ridge National Laboratory, Oak Ridge, TN 37831 USA

[2]Computational Urban Sciences Group, Oak Ridge National Laboratory, Oak Ridge, TN 37831 USA

[3]Materials Science and Technology Division, Oak Ridge National Laboratory, Oak Ridge, TN 37831 USA

CORRESPONDING AUTHOR: Xiao-Ying Yu (e-mail: yuxiaoying@ornl.gov).

**ABSTRACT** Access to continuous, quality assessed meteorological data is critical for understanding the climatology and atmospheric dynamics of a region. Research facilities like Oak Ridge National Laboratory (ORNL) rely on such data to assess site-specific climatology, model potential emissions, establish safety baselines, and prepare for emergency scenarios. To meet these needs, on-site towers at ORNL collect meteorological data at 15-minute and hourly intervals. However, data measurements from meteorological towers are affected by sensor sensitivity, degradation, lightning strikes, power fluctuations, glitching, and sensor failures, all of which can affect data quality. To address these challenges, we conducted a comprehensive quality assessment and processing of five years of meteorological data collected from ORNL at 15-minute intervals, including measurements of temperature, pressure, humidity, wind, and solar radiation. The time series of each variable was pre-processed and gap-filled using established meteorological data collection and cleaning techniques, i.e., the time series were subjected to structural standardization, data integrity testing, automated and manual outlier detection, and gap-filling. The data product and highly generalizable processing workflow developed in Python Jupyter notebooks are publicly accessible online. As a key contribution of this study, the evaluated 5-year data will be used to train atmospheric dispersion models that simulate dispersion dynamics across the complex ridge-and-valley topography of the Oak Ridge Reservation in East Tennessee.

## BACKGROUND

This dataset was developed with the goal of cleaning atmospheric data that can be used in dispersion models to predict the movement of airborne hazardous materials given an emergency event or site-planning needs. Atmospheric dispersion models are essential for emergency preparation and preparedness as well as for safety baseline development at National Laboratory facilities [1–5]. This includes the Oak Ridge National Laboratory (ORNL) and Y-12 Security Complex located in eastern Tennessee on the Oak Ridge Reserve (ORR). Historically, research activities at ORNL and Y-12 have produced and used materials at risk, and several ORR sites and structures are designated clean-up areas by the U.S. Environmental Protection Agency (EPA)'s Superfund program [6–8]. Thus, it is important to prepare for, identify, respond to, and mitigate the potential release of hazardous substances to protect the health of American citizens, as well as the ecosystems and resources that they depend on [1–5]. At ORR, Tennessee ridge and valley topography creates complex atmospheric flow and turbulence interactions that make atmospheric modeling of hazardous material difficult [9]. Simulating complex atmospheric motions and diffusion using computational models is often a data-informed process, which makes the quality of simulation results highly dependent on the underlying data [10–11]. Therefore, ensuring the accuracy and reliability of meteorological observations used to develop dispersion models is crucial. Usually, hourly data are already quality controlled, but to perform high-resolution dispersion modeling, higher resolution data are needed. Meteorological towers face power surges, sensor degradation over time, as well as glitching or failing sensors, all of which may affect data quality and coverage. High resolution (sub-hour) meteorological data can be especially difficult to quality control given that databases can be terabytes large, and sensors have variable acquisition frequencies due to instrument configurations.

To address this challenge, we aligned our data processing procedures with the standards and methods established by facilities and projects that maintain high-quality meteorological data measurements. For example, the National Climatic Data Center (NCDC) performs integrity checks for repeated and duplicated values, identifies potential outliers using a mixture of statistical thresholding methods, locates violations in multivariate relationships, and performs temporal tests for identifying spikes and dips [12]. Similarly, for the DOE's nuclear waste cleanup project at the Hanford Site, researchers developed a guide for preprocessing, cleaning, and assessing meteorology data for applications in applied risk and decision science [13]. They first applied statistical thresholding to remove outliers, dropped duplicate values, and then applied a moving average outlier detection method to remove temporal anomalies. We adapted the procedures from NCDC and DOE into three processing steps: (1) standardization of file structures and headers; (2) data quality control using threshold-based integrity testing, automated outlier detection via moving average filters, and manual outlier detection of false negative outliers; and (3) gap-filling with existing hourly data, and then linear interpolation.

The ORNL surface meteorology measurement dataset contains meteorological tower data collected by six towers located on the ORR from 2017 to 2022 as shown in **Figure 1**. We quality controlled and standardized 12 key variables that are useful in dispersion modeling, which includes temperature, pressure, precipitation, wind direction, lateral wind speed, vertical wind speed, peak wind speed, horizontal wind direction variability, wind direction variability, absolute humidity, relative humidity, and solar radiation. Not all towers collected the same suite of variables, and each tower collected data at one or more heights above ground level. Data collected at various sites above and below local ridgetops make for useful atmospheric profiling within ridge-and-valley topography. This data is useful not only for dispersion modeling, but also for East Tennessee weather, for the DOE complex, and for researchers interested in atmospheric science and use of long-term meteorological data in general.

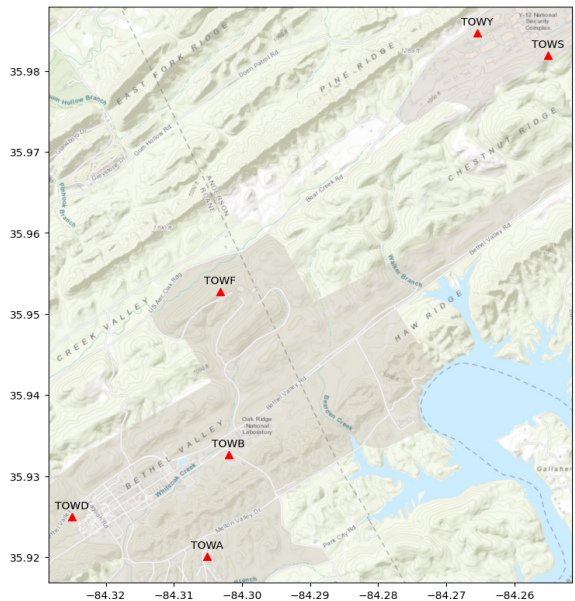

FIG. 1. Map of towers on ORNL and Y-12 campuses with a topographical basemap. The tower centroids are indicated using a red triangle.

**COLLECTION METHODS AND DESIGN**

**Data Collection**

Two datasets were curated and selected in this study: a not yet quality controlled (non-QC) quarter-hourly timeseries and a quality controlled (QC) hourly timeseries that were hosted on the ORNL's public website [14]. The observation site contains meteorological tower sensor data from 1998 to 2022. In general, ORNL tower sensor placement follows Environmental Protection Agency (EPA)'s guidance on meteorological monitoring for regulatory modeling applications [15]. Guidance states that wind data should be collected at, or above 10-m, though actual sensor placement may be higher due to terrain or physical obstructions. Temperature and humidity are measured at or above 2-m but less than or equal to 10-m. Precipitation and solar radiation are measured near the surface. Moreover, not all meteorological towers at ORR were built with regulatory modeling applications in mind, so some towers do not collect the same variables as other towers. Sensor height placement specifications are detailed in **Supplementary Table S1**.

We retrieved 2017–2022 archival data from Towers A, B, D, F, S, and Y, because they offer instrument rich and consistent data streams. Towers A, B, D, and F are located on the ORNL campus, with tower F situated on a local ridgetop to capture atmospheric dynamics at and above ridges (see **Figure 1**), which on campus are around 350-m above sea level. Towers S and Y are located on the Y-12 National Security Complex campus, with tower S also at ridgetop level. Data were collected by a total of 12 sensors among all 6 towers.

A system-based workflow was created within a Python Jupyter environment to standardize and quality control the data. The workflow, illustrated in **Figure 2**, is comprised of the following key components: (1) file standardization, (2) data quality control, and (3) data gap-filling.

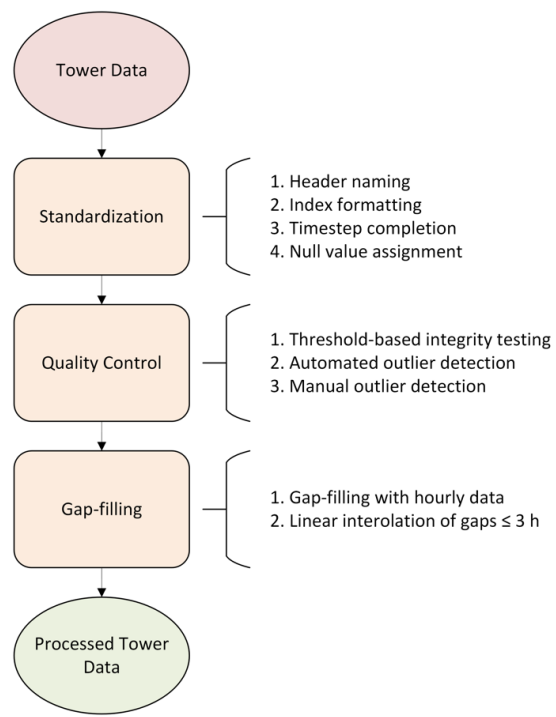

FIG. 2. A generalized workflow for processing meteorological timeseries data.

**Standardization**

We accessed time series data from 12 key meteorological variables described in **Table 2**. We preprocessed the 15-min data using a combination of manual analysis and automated procedures using Python Jupyter notebooks. The source archival data were divided into annual, monthly, and sub-monthly text files, all with disparate naming schemas that ultimately were merged into one annual file. First, tabular structures were standardized to have one row of header columns and datetime indices. The files were loaded into Pandas dataframes, of which we iteratively identified, merged, and standardized column names that contained misspellings, misleading units or titles, extra whitespace, and/or unnecessary characters. Additionally, null values were identified and standardized to -999, and the datetime index was localized to Coordinated Universal Time (UTC) with equal intervals between time steps.

TABLE 2. Variables, units, and a list of towers that collected data on each variable.

| Variable | Unit | Towers |
|---|---|---|
| Temperature | °C | A, B, D, F, S, Y |
| Pressure | mb | A, B, D, F, Y |
| Precipitation | in | A, B, D, F, Y |
| Wind Direction | degree | A, B, D, F, S, Y |
| Lateral Wind Speed | mph | A, B, D, F, S, Y |
| Peak Lateral Wind Speed | mph | A, B, D, F, S, Y |
| Vertical Wind Speed | mph | A, B, D, F, S, Y |
| Horizontal Wind Direction Variability ($\sigma\theta$) | degree | A, B, D, F, S, Y |
| Wind Direction Variability ($\sigma\varphi$) | degree | A, B, D, F, S, Y |
| Absolute Humidity | $g/m^3$ | A, D, F |
| Relative Humidity | percent | A, B, D, F, Y |
| Solar Radiation | $W/m^2$ | D, F, Y |

**Quality Control**

1) INTEGRITY TESTING

The first step of quality control is to ensure the basic integrity of the data by removing out-of-range, continuous, or rapidly fluxing values. Firstly, QC hourly data were resampled to 15-min intervals and interpolated using cubic splining. This resampling made it possible to use the QC hourly data for filling gaps in the non-QC quarter-hourly data later. Because the hourly data were already quality assessed and controlled, no further processing was required. Thus, the rest of this paper will focus on how the quarter-hourly data were quality controlled.

First, missing columns were calculated using other existing columns. For example, if a tower's Pandas dataframe only contained a column with temperature in Fahrenheit, a column with temperature in Celsius was computed. If a tower's dataframe was missing a particular variable (e.g., absolute humidity) that could be derived from others (e.g., pressure, temperature, and relative humidity), that missing variable was generated into a new column.

Next, the 12 variables of interest were selected for processing. Values outside of acceptable ranges were masked as outliers. The definition of an acceptable range was dependent on the variable measured. For example, temperature and barometric pressure limits were defined according to min/max QC hourly observations from 1998 to 2024. Relative humidity, absolute humidity, solar radiation, and wind speed ranges were based on their physical limits. For example, relative humidity values were capped between 0 and 100-percent because humidity cannot be negative nor can it be over 100, i.e., the maximum amount of water air can hold before condensation. Lastly, we dropped temporally unchanging values lasting greater than 6 hours for all variables except precipitation, relative humidity, and solar radiation; these variables have periods of constant values reflecting no rain, rain, and nighttime conditions, respectively. See **Supplementary Table S2** for more details on quarter-hourly thresholding information.

2) AUTOMATED OUTLIER DETECTION

A simple moving average filter was employed to detect and remove values beyond a z-score threshold. For each variable, two moving average-based outlier detection methods were tested. For the first method, referred to as the single-tower approach, an individual tower and sensor height were selected. For each variable at the sensor height, the moving average was calculated along with the standard deviation of the variable's time series. For a given time point $t$, the simple moving average, $SMA$, is calculated over a window of size $w$. Equation (1) for the single tower approach's moving average is

$$SMA_t = \frac{1}{w}\sum_{i=0}^{w-1} X_{t-i}. \qquad (1)$$

Equation (2) for standard deviation $\sigma_t$ of the values within the rolling window is defined as

$$\sigma_t = \sqrt{\frac{1}{w}\sum_{i=0}^{w-1}(x_{t-1} - SMA_t)^2}. \qquad (2)$$

An observation $x_t$ is considered an outlier if the absolute difference between the observation and moving average exceeds a defined z-score threshold $k$, which is a multiple of the standard deviation.

However, the single-tower approach can be improved by leveraging data from nearby towers at similar sensor heights, or the multi-tower approach. For each tower variable observation $X_t^{(i)}$, a moving average of window size $w$ is applied. Equation (3) for that moving window is,

$$SMA_t^{(i)} = \frac{1}{w}\sum_{j=0}^{w-1} X_{t-j}^{(i)}. \qquad (3)$$

Then, for towers $G$ that are grouped by their similar altitudes, we compute Equation (4) of the group's variable $GM$ at time $t$ as the following,

$$GM_t = \frac{1}{|G|}\sum_{i \in G} X_t^{(i)}. \qquad (4)$$

Lastly, we calculate the $GM$ moving average as $SMA_t^G$. A variable observation $X_t^{(i)}$ is considered a true outlier if the absolute difference between the observation $X_t^{(i)}$ and moving average $SMA_t^{(i)}$ exceeds a defined z-score threshold $k$, and the absolute difference between the group's observation $GM_t$ and moving average $SMA_t^G$ exceeds a defined z-score threshold $k$. Equation (5) is used,

$$\left(|X_t^{(i)} - SMA_t^{(i)}| > k \cdot \sigma_t^{(i)}\right) \text{ AND } \left(|GM_t - SMA_t^G| > k \cdot \sigma_t^G\right). \qquad (5)$$

Where $k$ is a predefined z-score threshold, and $\sigma_t^{(i)}$ and $\sigma_t^G$ represent the standard deviations of the individual tower and the group mean, respectively. To optimize time and effort, we visually assessed the presence and frequency of outliers to establish a baseline goal for removal. For each variable, all combinations of window sizes $w$ (i.e., 1, 3, 6, and 12 h.;

1, 3, 6, and 12 days; 30, 60, 90, and 120 days) and z-score thresholds $k$ (i.e., 3 and 4 standard deviations) were evaluated. The best combination of window size $w$ and z-score threshold $k$ was determined by comparing the count, date, and time of identified outliers to those estimated in the baseline goal.

Once we identified the ideal window size and threshold combination, the accuracy of the single-tower and multi-tower approaches were assessed by selecting 20% of the detected outliers and manually checking if they were true or false positives. To measure accuracy, we divided the number of true positive outliers by the sample population $n$ of all detected outliers. We found that the multi-tower approach yielded higher accuracy than the single-tower approach for all variables, but accuracy among towers and variables varied. Temperature outlier detection reached 83% accuracy, while wind speed reached only 22% accuracy. In such instances where accuracy was low, a larger threshold was employed to remove less data.

3) MANUAL OUTLIER DETECTION

After outliers were automatically removed for each variable, remaining false negatives were manually identified and dropped. We also manually removed outliers for horizontal wind direction variability, wind direction variability, and wind direction; primarily because these variables were derived from cyclical variables unfit for linear outlier detection methods like moving windows. In the interest of time, meteorological data experts manually and subjectively addressed the cyclical data to identify visually apparent outliers. For all variables, dropped and no-data values were filled with -999. The final outlier detection method(s), moving window size $w$, and z-score threshold $k$ selected for each variable are shown in **Table 3**.

*TABLE 3. Summary of outlier detection method(s), window sizes, and z-score thresholds for each variable.*

| Variable | Outlier Detection Method | Size $w$ | Threshold $k$ |
|---|---|---|---|
| Temperature | Multi-Tower + Manual | 1 day | 3 |
| Absolute Humidity | Multi-Tower + Manual | 6 hours | 3 |
| Relative Humidity | Multi-Tower + Manual | 12 hours | 4 |
| Wind Speed | Multi-Tower + Manual | 3 hours | 3 |
| Peak Wind Speed | Multi-Tower + Manual | 6 hours | 4 |
| Horizontal Wind Direction Variability (σθ) | Drop 0 + Manual | n/a | n/a |
| Wind Direction Variability (σφ) | Drop 0 + Manual | n/a | n/a |
| Vertical Wind Speed | Manual | n/a | n/a |
| Barometric Pressure | Multi-Tower + Manual | 3 hours | 3 |
| Solar Radiation | Multi-Tower + Manual | 12 hours | 4 |

**Gap-filling**

To create a gap-filled product, we first filled gaps with hourly data that was interpolated to 15-min time steps using the cubic spline method. The hourly data was complete and already quality controlled. To fill any remaining small gaps ($\leq 3$ h) in a variable's time series, we first added a 3-hour buffer to the beginning and end of the time series to prevent interpolation errors at the data's temporal boundaries. To prevent spikes or dips in interpolated data, we created a smoothed version of the time series using the Savitzky-Golay method configured with a 3-hour window length, poly-order of 2, and the nearest neighbor method. Then, we used linear interpolation to fill small gaps within a 3-hour or smaller window. At the edge of large gaps, outliers may have been created during the interpolation process, so we applied a global and local clip to lessen the magnitude of newly generated outliers. The remaining small gaps were filled with this smoothed, interpolated, and clipped time series. An example showing missing data before and after gap-filling is shown in **Figure 3**.

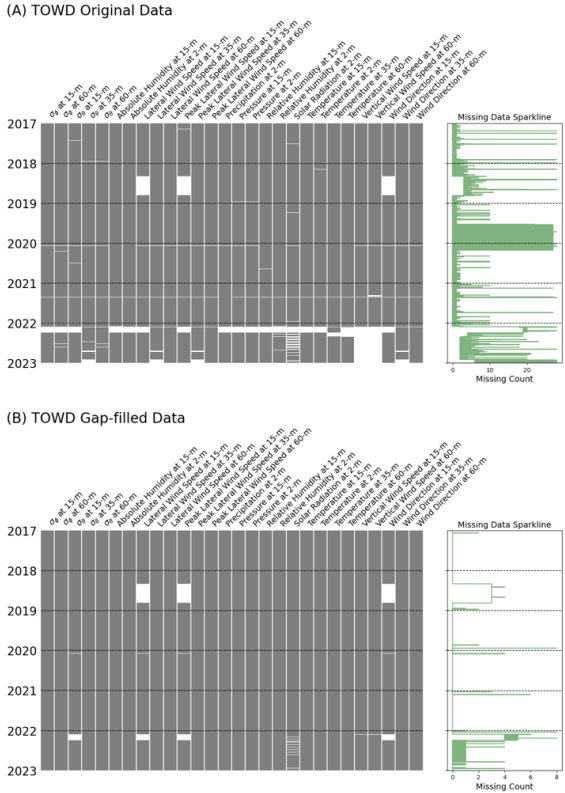

FIG. 3. An MSNO graph of missing data in Tower D's five-year time series (A) before and (B) after gap-filling. Data are represented by each gray bar, with white breaks representing data gaps. In green, the missing data sparkline displays the count of variables with missing data at each 15-min time stamp from 2017 to 2022.

**VALIDATION AND QUALITY**

To ensure the quality of the data, we applied extensive preprocessing techniques by standardizing table headers, data types, and no-data representatives. We removed erroneous data using simple thresholding, identifying temporally continuous values, and applying both automated and manual outlier detection techniques. We filled gaps in data by calculating missing variables, filling gaps with quality-controlled hourly data, and filling gaps less than 3 h using interpolation techniques, all of which were described in the Collection Methods and Design section. The result of our pre-processing and quality control workflow are exemplified in **Figure 4**, which displays "original" non-QC quarter-hourly temperature measurements versus quality-controlled, gap-filled quarter-hourly temperature measurements for Tower D at the 15-m sensor height. More meteorological variables and time series plots for Tower D sampled at random years, months, and days, can be found in **Supplemental Figures S1–3**.

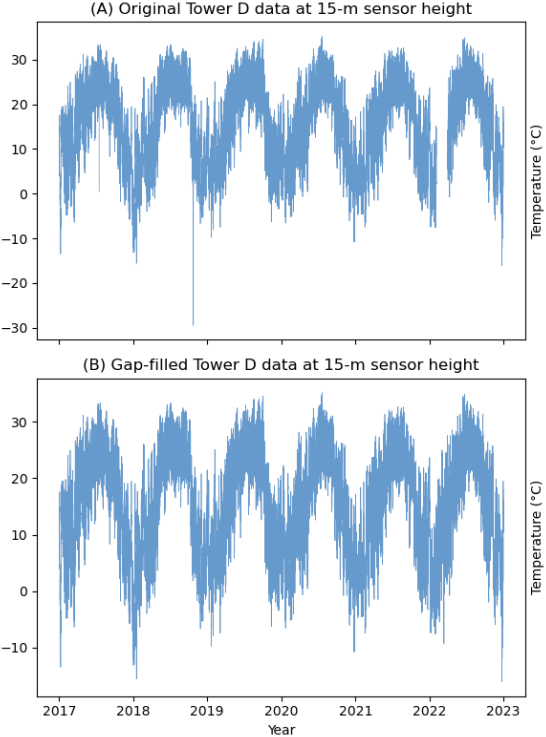

FIG. 4. A five-year time series comparison of (A) pre-quality controlled and (B) quality controlled and gap-filled temperature at 15-m sensor height of Tower D.

To get the highest-quality interpolated data, we first used QC hourly data to fill gaps. Then, we tested two common methods for interpolating small gaps ≤ 3 h in a time series, namely linear interpolation and the Piecewise Cubic Hermite Interpolating Polynomial (PCHIP) method. Both were computationally efficient, but PCHIP is known to preserve the shape and monotonicity of time series curves, ideally preventing erroneous oscillations or overshoots during interpolation [16]. As shown in **Table 4**, the linear interpolation gap-filling method was marginally more accurate than PCHIP, with an average Root-Mean Square Error (RMSE) of 10.42 (10.60). Wind direction, solar radiation, and wind direction variabilities were not interpolated as well as other variables because of their partial or full cyclical nature and rapidity of change over time. For example, solar radiation has unique, sudden jumps due to cloud cover, as well as a strong diurnal cycle that is not well suited for linear interpolation methods.

*TABLE 4. Mean and median RMSE for linear vs. PCHIP interpolation of all heights and towers at ORR.*

|  |  | PCHIP | | Linear | |
| --- | --- | --- | --- | --- | --- |
| Variable | Unit | Mean | Median | Mean | Median |
| Absolute Humidity | g/m$^3$ | 0.18 | 0.18 | 0.18 | 0.18 |
| Barometric Pressure | mb | 0.14 | 0.15 | 0.14 | 0.15 |
| Peak Wind Speed | mph | 2.03 | 1.95 | 1.99 | 1.93 |
| Precipitation | in | 0.01 | 0.01 | 0.01 | 0.01 |
| Relative Humidity | percent | 1.17 | 1.16 | 1.25 | 1.23 |
| Horizontal Wind Direction Variability ($\sigma\theta$) | degree | 14.5 | 14.9 | 14.16 | 14.56 |
| Wind Direction Variability ($\sigma\varphi$) | degree | 3.62 | 3.67 | 3.52 | 3.57 |
| Solar Radiation | W/m$^2$ | 56.93 | 56.6 | 56.22 | 55.86 |
| Temperature | °C | 0.19 | 0.19 | 0.19 | 0.19 |
| Vertical Wind Speed | mph | 0.26 | 0.25 | 0.26 | 0.25 |
| Wind Direction | degree | 58.81 | 56.28 | 57.88 | 55.45 |
| Wind Speed | mph | 0.93 | 0.93 | 0.92 | 0.92 |

**RECORDS AND STORAGE**

Data are publicly accessible on the ORNL meteorology webpage [14]. As shown in **Table 5,** the data are utf-8-sig encoded into CSV files and are stored in seven directories representing each processing step: (1) hourly-qc, (2) original-qc, (3) manual-outlier-id, (4) final-qc, (5) gapfilled-qc, (6) gapfilled-bool, and (7) supplementary. We provided data at each stage for transparency, easier troubleshooting and error identification, and for users to clearly understand how the data changed between each step. In each directory, there is one file per tower, where towers are denoted as TOWA, TOWB, TOWD, TOWF, TOWS, TOWY. Directory 1, hourly-qc, contains QC hourly data interpolated into 15-minute time steps. Directory 2, original-qc, contains non-QC quarter-hourly data with standardized column headers, merged duplicate columns, corrected indices, and a quick look-over for obvious structural or syntax errors. Directory 3, manual-outlier-id, contains the indices of manually identified outliers missed by automated outlier removal. Directory 4, final-qc, contains data cleaned using both the automated and manual outlier removal processes. Directory 5, gapfilled-qc, contains the same data as final-qc but with gaps filled. Directory 6, gapfilled-bool, contains a Boolean-masked table that can be used to see which values in gapfilled-qc were filled with hourly data or interpolated.

We also include a directory containing supplementary data files: met_inst_ranges, met_towers_info, and metadata, which provide acceptable data ranges for each meteorological variable, tower-level metadata, and file-level metadata, respectively. Most CSV file columns correspond to a meteorological variable's short name, which are formatted as *VariableUnit_Height*. The translation from short name to long name, as well as a description and unit, are included in the file-level metadata. Each row of the table represents a 15-min interval with an associated timestamp in UTC.

*TABLE 5. Qualified data file structure for the ORR.*

|  | Processing Stage | Directory Name | CSV File(s) |
| --- | --- | --- | --- |
| 1 | hourly-qc | met_towers_ 2017-2022_ hourly-qc | TOWA, TOWB, TOWD, TOWF, TOWS, TOWY |
| 2 | original-qc | met_towers_ 2017-2022_ original-qc | TOWA, TOWB, TOWD, TOWF, TOWS, TOWY |
| 3 | manual-outlier-id | met_towers_ 2017-2022_ manual-outlier-id | TOWA, TOWB, TOWD, TOWF, TOWS, TOWY |
| 4 | final-qc | met_towers_ 2017-2022_ final-qc | TOWA, TOWB, TOWD, TOWF, TOWS, TOWY |
| 5 | gapfilled-qc | met_towers_2017-2022_gapfilled-qc | TOWA, TOWB, TOWD, TOWF, TOWS, TOWY |
| 6 | gapfilled-bool | met_towers_2017-2022_gapfilled-bool | TOWA, TOWB, TOWD, TOWF, TOWS, TOWY |
| 7 | supplementary | | met_inst_ranges |
| 7 | supplementary | | met_towers_info |
| 7 | supplementary | | metadata |

## INSIGHTS AND NOTES

This site-specific dataset was curated for use in atmospheric dispersion modeling at and around ORR. The processed meteorological data will also be used for an in-depth analysis of weather and climate history at the laboratory, which is essential background material needed for assessments based on dispersion modeling, as well as other safety and risk analyses using applied mathematical methods. Further, this dataset provides a useful guide for the thorough data cleaning and gap-filling of long-term meteorological variables collected in a unique, mid-latitude, ridge-and-valley environment like the ORR, presenting a valuable source of data for studying atmospheric dynamics in highly complex terrain. Further, this work presents a robust framework for ensuring data quality through mathematical and statistical approaches, resulting in benchmark datasets for future research.

## SOURCE CODE AND SCRIPTS

The Python notebooks used to process the data, as well as some CSV data, are publicly archived on Github: https://github.com/msteckle/orr_met_data_processing.

## ACKNOWLEDGMENTS AND INTERESTS

M. R. Steckler and K. R. Birdwell curated and analyzed the data. M. R. Steckler, K. R. Birdwell, and X.-Y. Yu wrote the manuscript. X.-Y. Yu secured funding and supervised the project. All authors reviewed the manuscript.

Research efforts were supported by the Nuclear Safety Research and Development (NSR&D) program sponsored by National Nuclear Security Administration (NNSA) Office of Environment, Health, Safety and Security (EHSS) of the U.S. Department of Energy (DOE).

This manuscript has been authored by UT-Battelle, LLC, under contract DE-AC05-00OR22725 with the US DOE. The US government retains and the publisher, by accepting the article for publication, acknowledges that the US government retains a nonexclusive, paid-up, irrevocable, worldwide license to publish or reproduce the published form of this manuscript, or allow others to do so, for US government purposes. DOE will provide public access to these results of federally sponsored research in accordance with the DOE Public Access Plan (https://www.energy.gov/doe-public-access-plan).

The article authors have declared no conflicts of interest.

Supplementary information

# Descriptor:
# Five years of meteorological surface data at Oak Ridge Reserve in Tennessee

MORGAN R. STECKLER[1], KEVIN R. BIRDWELL[1], HAOWEN XU[2], AND XIAO-YING YU[3*]

[1]Integrated Computational Earth Sciences Group, Oak Ridge National Laboratory, Oak Ridge, TN 37831 USA
[2]Computational Urban Sciences Group, Oak Ridge National Laboratory, Oak Ridge, TN 37831 USA
[3]Materials Science and Technology Division, Oak Ridge National Laboratory, Oak Ridge, TN 37831 USA

CORRESPONDING AUTHOR: Xiao-Ying Yu (e-mail: yuxiaoying@ornl.gov).

## Table of Contents

Supplementary information

Additional information was provided to support the main text, including tables with detailed specifications of each tower and variable-specific thresholding ranges, as well as example figures containing time series plots of quality assessed data.

**Supplemental Tables**

**Table S1. Sensor and tower specifications at the ORR.**

| Tower | Tower Base Altitude (m) | Sensor Height from Base (m) | Sensor Altitude (m) |
|---|---|---|---|
| Tower A | 263 | 15 | 278 |
| | | 30 | 293 |
| Tower B | 255 | 15 | 270 |
| | | 30 | 285 |
| Tower D | 261 | 2 | 263 |
| | | 15 | 276 |
| | | 35 | 296 |
| | | 60 | 321 |
| Tower F | 354 | 10 | 364 |
| Tower S | 352 | 25 | 377 |
| Tower Y | 290 | 15 | 305 |
| | | 33 | 323 |

Supplementary information

**Table S2. Basic integrity testing thresholds for each variable.**

| Variable | Error Type | Units | Min | Max | Min Description | Max Description |
|---|---|---|---|---|---|---|
| TempF | Out of Range | °F | -30 | 110 | Sensor limit | Sensor limit |
| TempF | Hourly Rate of Change | °F | | 10 | | ORR maximum rate of change observed (1998-2024) |
| RelHum | Out of Range | % | 0 | 100 | Physical limit | Physical limit |
| RelHum | Hourly Rate of Change | % | | 25 | | ORR maximum rate of change observed (1998-2024) |
| AbsHum | Out of Range | mg/m$^3$ | 0.1 | 28 | Physical limit | ORR maximum observed (1998-2024) |
| BarPresMb | Out of Range | mb | 965.1 | 1032.8 | ORR minimum observed (1998-2024) | ORR maximum observed (1998-2024) |
| SolarRadWm2 | Out of Range | W/m$^2$ | 0 | 1250 | Physical limit | ORR latitude-based physical limit |
| WDir | Out of Range | degree | 0 | 360 | Minimum compass value | Maximum compass value |
| WSpdMph | Out of Range | mph | 0 | 50 | Physical limit | Above ORR maximum value observed (1998-2024) |
| PkWSpdMph | Out of Range | mph | 0 | 115 | Physical limit | Sensor limit |
| VSSpdMph | Out of Range | mph | -3 | 3 | Determined to be minimum credible observed based on multiple tower obs (2024-2024) | Determined to be maximum credible observed based on multiple tower obs (2024-2024) |
| Sigma (Theta) | Out of Range | degree | 0 | 180 | Minimum compass value | Maximum value away from 0 or 360 |
| SigPhi | Out of Range | degree | -60 | 60 | Sensor limit | Sensor limit |
| PrecipIn | Out of Range | in | 0 | 8 | Physical limit | Above ORR maximum value observed (1947-2024) |

Supplementary information

## Supplemental Figures

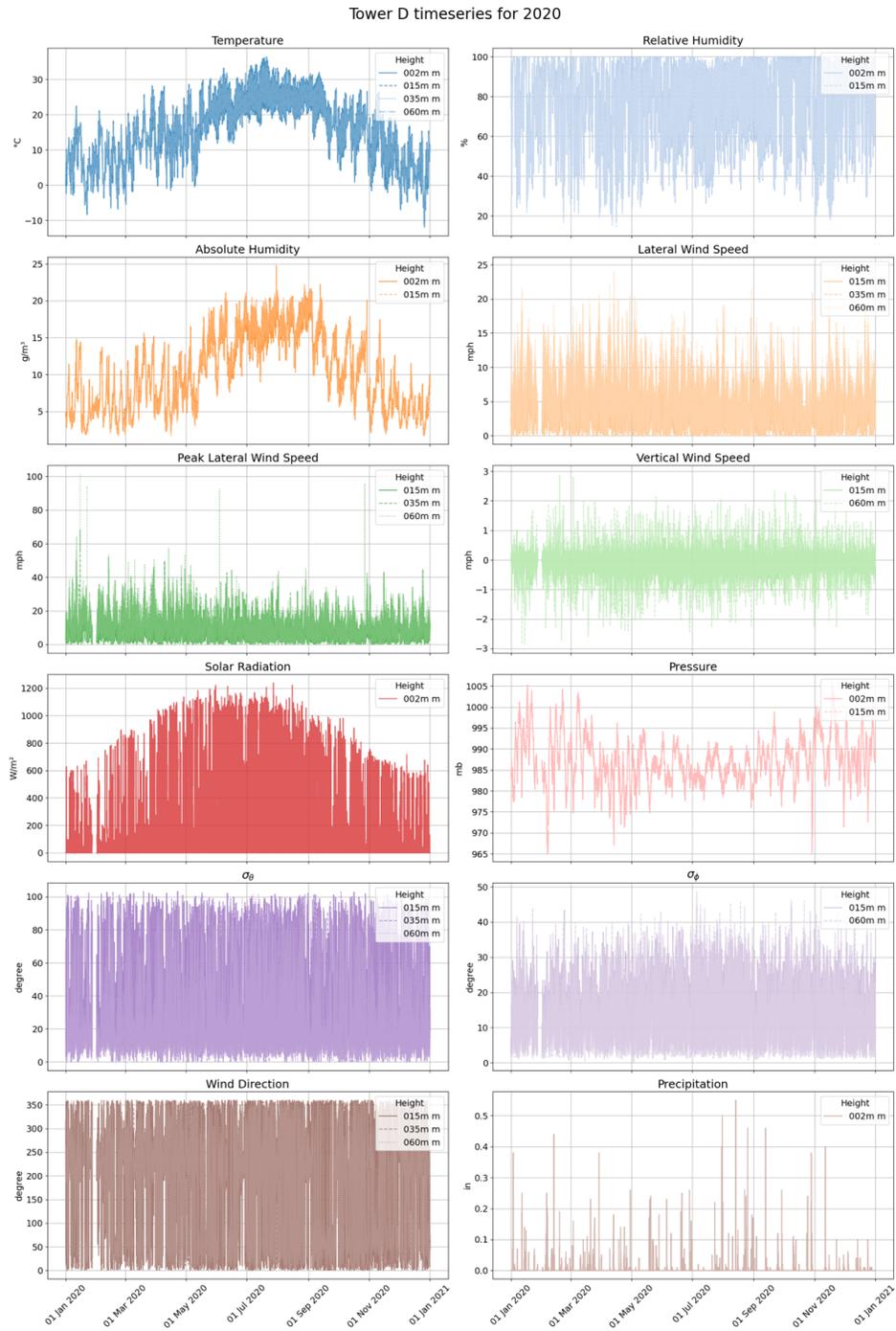

**Figure S1. Example of quality-controlled quarter-hourly data in 2020 for Tower D.**

Supplementary information

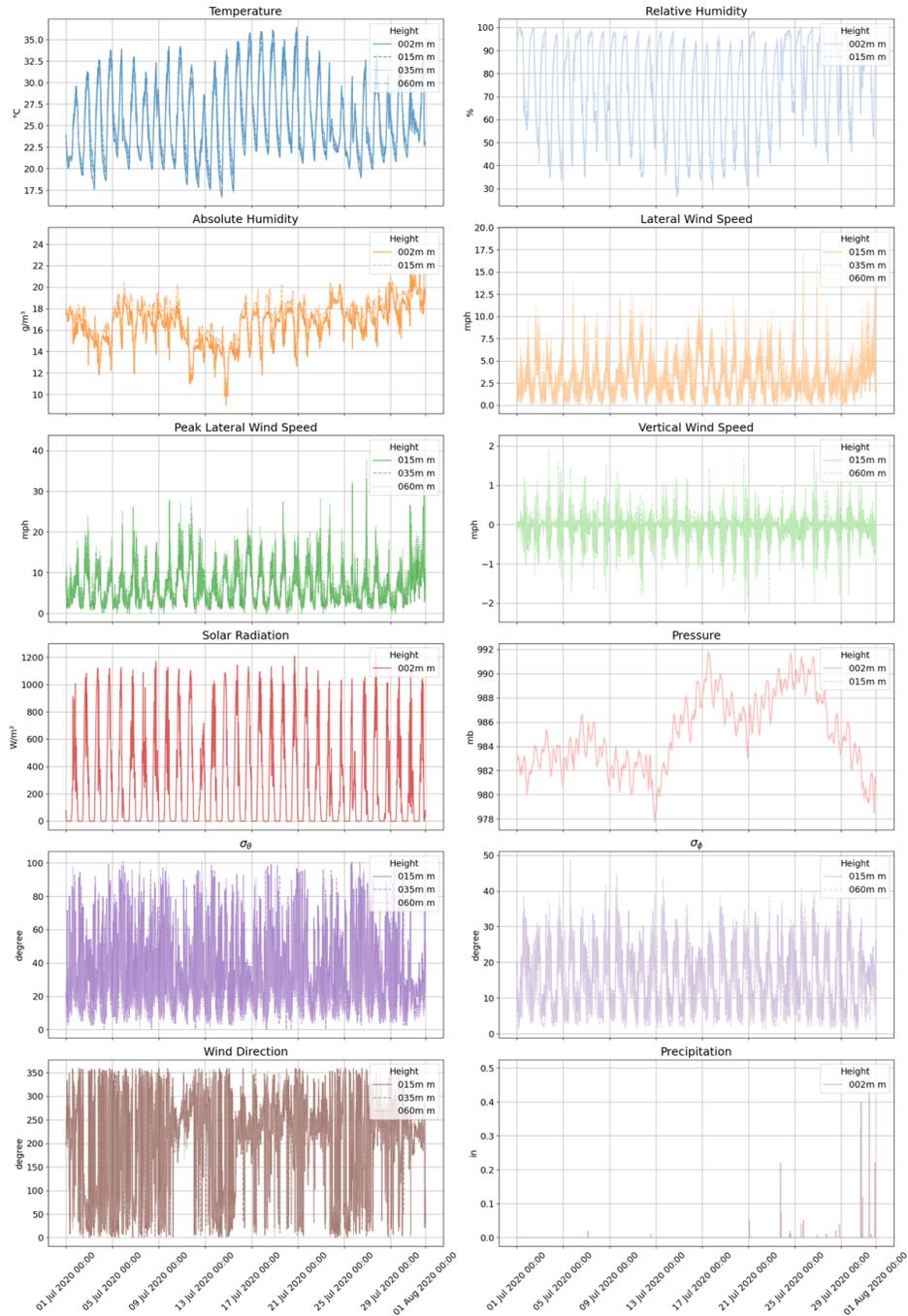

**Figure S2. Example of quality-controlled quarter-hourly data in July 2020 for Tower D.**

Supplementary information

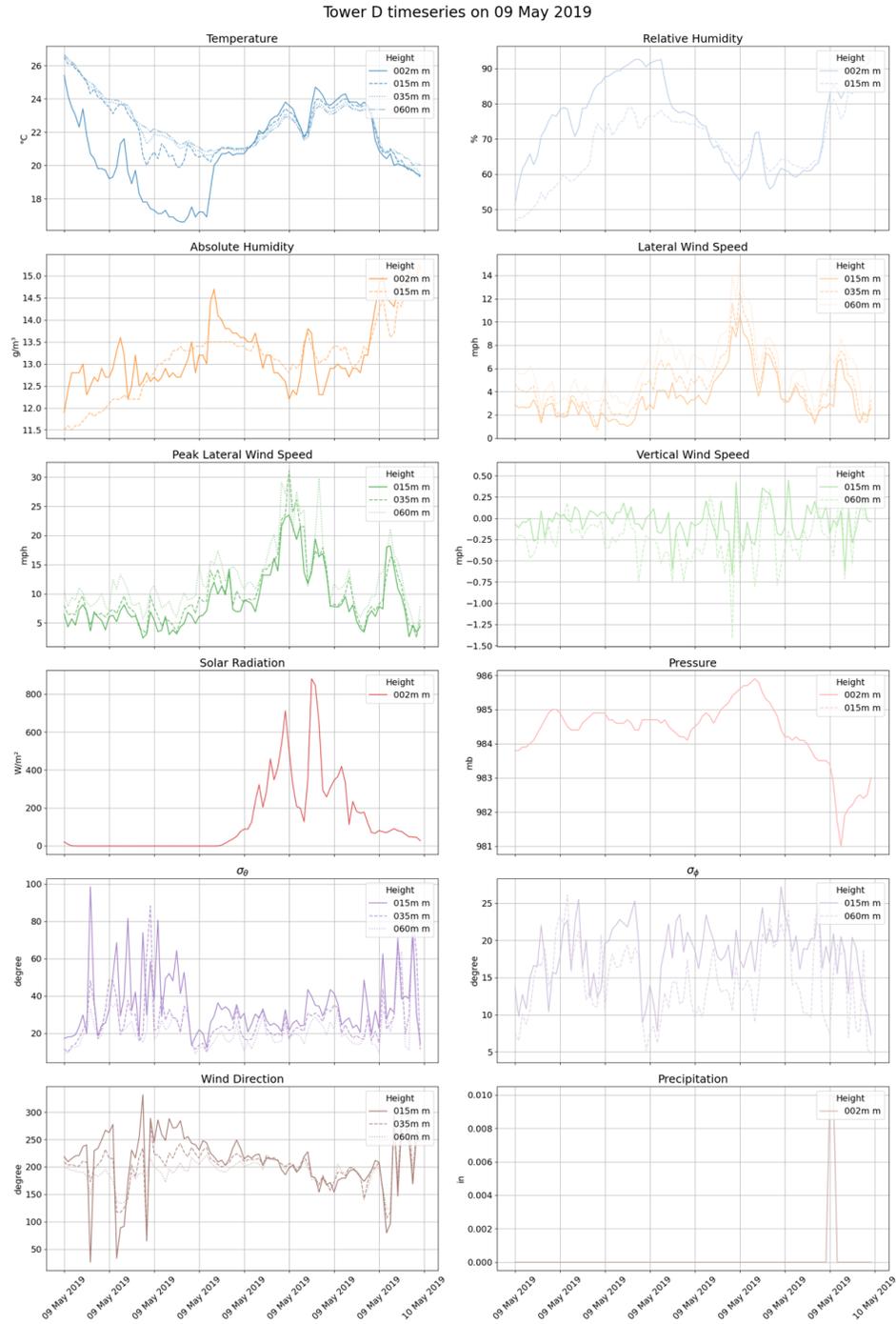

**Figure S3. Example of quality-controlled quarter-hourly data on 9 May 2019 for Tower D.**